\documentclass[letterpaper, 10 pt, conference]{ieeeconf}  

\IEEEoverridecommandlockouts                              

\usepackage{graphics} 
\usepackage{epsfig} 
\usepackage{mathptmx} 
\usepackage{times} 
\usepackage{amsmath} 
\usepackage{amssymb}  
\usepackage{booktabs}
\usepackage{cite}
\usepackage{threeparttable}
\usepackage{multirow}
\usepackage{color} 
\usepackage{tabularx}

\title{\LARGE \bf
Enhancing Bronchoscopy Depth Estimation through Synthetic-to-Real Domain Adaptation}

\author{Qingyao Tian, Huai Liao, Xinyan Huang, Lujie Li and Hongbin Liu
\thanks{Qingyao Tian is with Institute of Automation, Chinese Academy of Sciences, Beijing 100190, China, and with the School of Artificial Intelligence, University of Chinese Academy of Sciences, Beijing 100049, China.}%
\thanks{
Huai Liao, M.D. and Xin-yan Huang, M.D. are with Department of Pulmonary and Critical Care Medicine, The First Affiliated Hospital of Sun Yat-sen University, Guangzhou, Guangdong, China.
}
\thanks{
Lujie Li, M.D. is with Department of Radiology, The First Affiliated Hospital, Sun Yat-sen University, Guangzhou, Guangdong, China
}
\thanks{Hongbin Liu is with the Institute of Automation, Chinese Academy of Sciences, Beijing, 100190, China, and also with the School of Biomedical Engineering \& Imaging Sciences, King’s college London, London, SE1 7EU, UK. Corresponding author: Hongbin Liu (e-mail: liuhongbin@ia.ac.cn)}%
}

\begin{document}

\maketitle
\pubid{\begin{minipage}{\textwidth}\ \\[30pt] \centering
	This work has been accepted by ICRA 2024 Workshop on C4SR+.
\end{minipage}}

\begin{abstract}
Monocular depth estimation has shown promise in general imaging tasks, aiding in localization and 3D reconstruction. While effective in various domains, its application to bronchoscopic images is hindered by the lack of labeled data, challenging the use of supervised learning methods. In this work, we propose a transfer learning framework that leverages synthetic data with depth labels for training and adapts domain knowledge for accurate depth estimation in real bronchoscope data. Our network demonstrates improved depth prediction on real footage using domain adaptation compared to training solely on synthetic data, validating our approach.

\pubid{\begin{minipage}{\textwidth}\ \\[30pt] \centering
	This work has been accepted by ICRA 2024 Workshop on C4SR+.
\end{minipage}}

\end{abstract}

\section{INTRODUCTION}
Depth information plays a crucial role in localization and visual odometry \cite{teed2024deep}, as it provides essential 3D information about the environment. RGB-D cameras, which simultaneously capture color and depth, are widely used in general environments for depth acquisition. However, the compact size of bronchoscopes restricts the direct implementation of RGB-D cameras for depth measurement. Despite the progress in monocular depth estimation for general images \cite{bhat2023zoedepth}, applying these techniques to bronchoscopic images remains challenging because of the scarcity of labeled data for training of learning-based approaches.

To overcome these challenges, researchers have explored specific depth estimation methods suitable for bronchoscopic images \cite{karaoglu2021adversarial,tian2024ddvnb}. These methods predominantly utilize cycle adversarial architectures \cite{cyclegan} for unsupervised training, leveraging unpaired bronchoscopic frames and depth information from virtual bronchoscopy. However, the training process of GAN suffer from mode collapse and instability \cite{bau2019seeing}.

This work is motivated by the fact that although ground truth depth labels are inaccessible in real clinical settings, they are easily available in virtual bronchoscopy. In this work, we develop a domain adversarial network that is trained on labeled synthetic data, and transfer to real bronchoscopic data with high accuracy. Experiments highlight the effectiveness of transfer learning in leveraging synthetic data for depth estimation in bronchoscopy, demonstrating its potential for clinical applications.

\begin{figure}
    \centering
    \includegraphics[width=8.5cm]{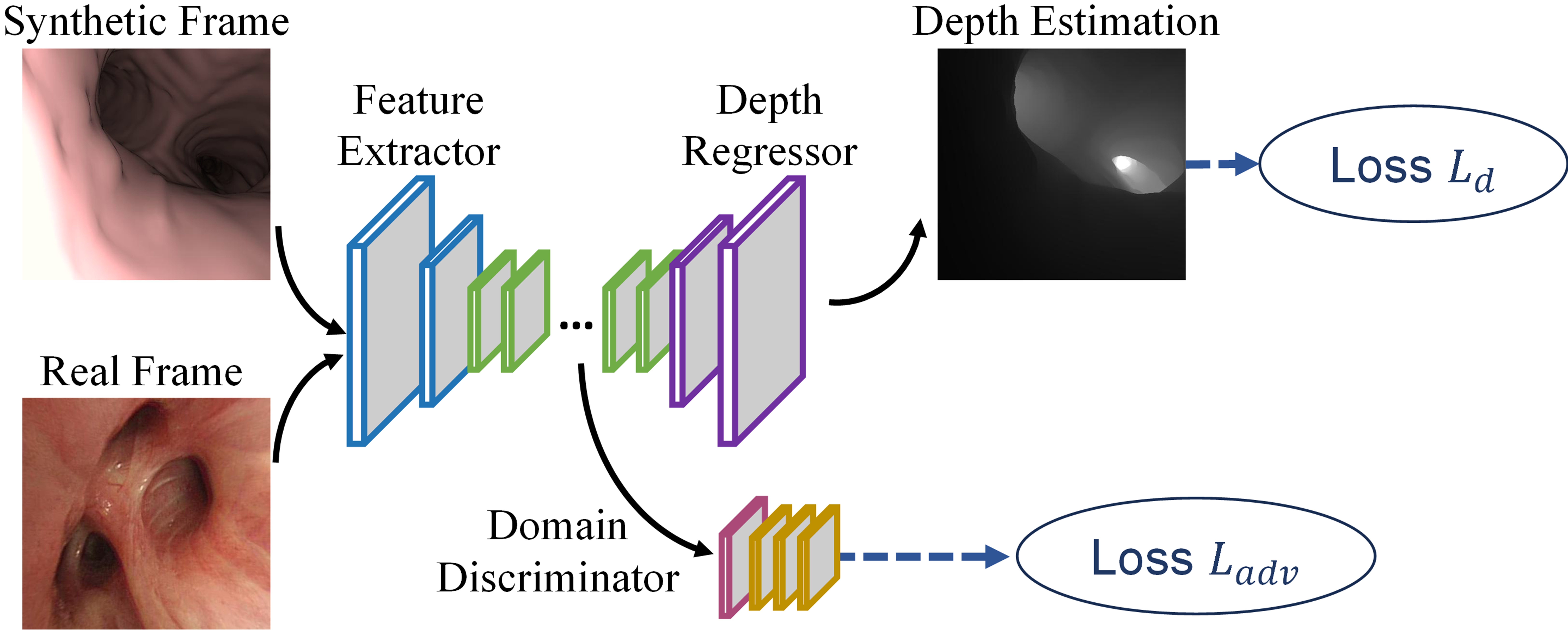}
    \caption{Diagram of the Synthetic-to-Real Domain Adaptation pipeline.}
    \label{figl}
\end{figure}

\section{METHOD}
The training diagram of proposed depth estimation network is illustrated in Fig. \ref{figl}. Inspired by domain adaptive neural network (DANN) \cite{ganin2016domain}, the framework comprises three key components: a feature extractor, a depth regressor, and a domain discriminator. The domain discriminator is trained simultaneously with the depth estimation network, aiming to differentiate between the source domain (\textit{e.g.}, synthetic data) and the target domain (\textit{e.g.}, real data). It processes the feature representation from the feature extractor and yields a probability that indicates the domain of the input data. The training objective is to enhance the domain discriminator's ability to distinguish between domains while concurrently training the feature extractor to reduce the discriminator's accuracy, resulting in features that are invariant across domains. Consequently, the depth estimation network is trained to learn features that are invariant across synthetic and real bronchoscopic data, facilitating effective domain adaptation.

In our depth estimation network, the overall training objective incorporates two pivotal loss for source domain (e.g., synthetic data) and adaptation to target domain (e.g., real-world data) respectively: 

\textbf{Source Domain Depth Estimation Loss} ($L_{\text{d}}$): This loss is articulated for the source domain, leveraging a blend of scale-and-shift invariant (SSI) loss and L1 loss for supervising the depth estimation. It is expressed as:


\begin{equation}
    L_{\text{d}} = \alpha \cdot L_{\text{SSI}} + \beta \cdot L_{\text{L1}},
\end{equation}
where $\alpha$ and $\beta$ denote the weighting coefficients for the SSI and L1 losses, respectively.

\begin{figure*}[tbp]
    \centering
    \includegraphics[width=13cm]{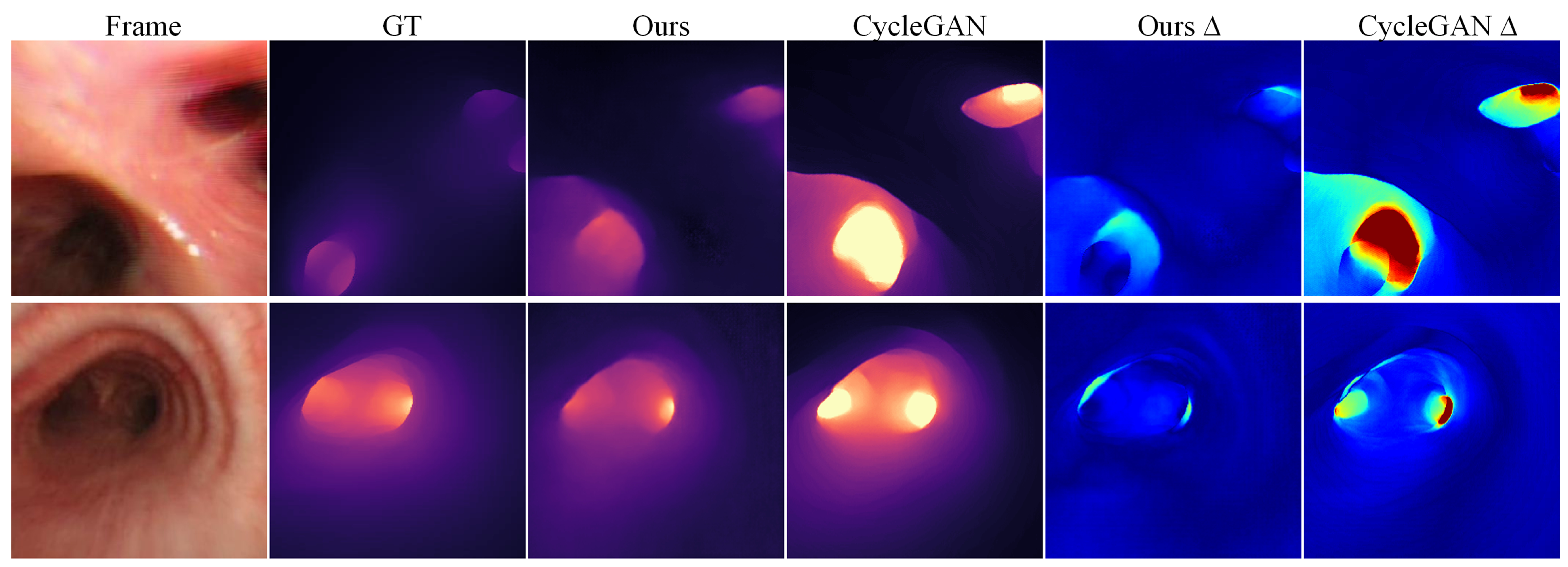}
    \caption{\textbf{Quantitative Evaluation on Real Bronchoscopy Data.} $\bigtriangleup$ denotes absolute error ranging from lowest (blue) to highest (red).}
    \label{fig2}
\end{figure*}

\textbf{Domain Adaptation Loss} ($L_{\text{adv}}$): To enable effective domain transfer from synthetic to real bronchoscopic imagery, we adopt the domain adversarial loss as per the DANN framework. This loss aims to render the feature representations domain-invariant by deceiving the domain classifier and is defined as:
\begin{equation}
L_{\text{adv}} = -\mathbb{E}_{x \sim p_{\text{source}}}\left[\log D(x)\right] - \mathbb{E}_{x \sim p_{\text{target}}}\left[\log (1 - D(x))\right],
\end{equation}

\noindent with $D$ being the domain classifier, and $x$ the input data.

Hence, the aggregate loss for our network's training is the summation of the aforementioned components:
\begin{equation}
    L_{\text{total}} = L_{\text{d}} + \gamma L_{\text{adv}},
\end{equation}

\noindent where $\gamma$ is weight term for domain adaption loss.

This structured loss function enable the network to learn depth estimations from synthetic data while adeptly adapting to the real bronchoscopic domain without the need for labeled depth in the real dataset.

\section{RESULTS}
\begin{table}[!tbp]
  \centering
  \caption{Depth Estimation Results on Real Bronchoscopy Data.}
    \begin{tabular}{@{}c@{\hspace{7pt}}c@{\hspace{7pt}}c@{\hspace{7pt}}c@{\hspace{7pt}}c@{}}
    \hline
    Method & SSIM $\uparrow$ & MAE (mm) $\downarrow$  & RMSE (mm) $\downarrow$ & $\delta_1$ \textless 1.25 $\uparrow$  \\
    \hline
    CycleGAN & 0.913  & 3.397 ± 1.885 & 5.566 ± 3.452 & 0.482  \\
    Ours \textit{w/o DA} & 0.931  & 2.813 ± 0.849 & 4.408 ± 1.297 & 0.498  \\
    Ours  & \textbf{0.932 } & \textbf{2.785 ± 0.849} & \textbf{4.382 ± 1.304} & \textbf{0.501} \\
    \hline
    \end{tabular}%
  \label{tab1}%
\end{table}%

The depth estimation network is trained using the PyTorch framework on an NVIDIA RTX3090 GPU. It comprises initial convolutional block, downsampling, nine residual blocks, and upsampling layers. A three-layer MLP with 1024 hidden units serves as the domain discriminator, processing features from the network's bottleneck, which are average-pooled before input. Initially trained on 22,570 synthetic frames from eight patient-derived airway models using Adam optimization (batch size 8, learning rate 1e-4, for 100 epochs), the network incorporates early stopping based on a validation set from an additional patient model. Following this, it is fine-tuned for 35 epochs using our domain adaptation method with hyperparameters $\alpha=1$, $\beta=100$, and $\gamma=0.1$.

Experiments were conducted to validate the proposed Synthetic-to-Real Domain Adaptation pipeline for depth estimation of real bronchoscopic images. We compared our domain-adapted depth estimation with CycleGAN \cite{cyclegan}, which is the baseline for most existing methods \cite{karaoglu2021adversarial,tian2024ddvnb}. We also compare with the same depth network architecture trained solely on virtual data. The validation was conducted on 2 patient cases, comprising 2457 frames. Ground truth data were labeled by manually registering virtual bronchoscopy frames with real footage and obtaining depth information in the simulator. Results are median-scaled before evaluation, as shown in Table \ref{tab1}. Additionally, we provide qualitative results in Fig. \ref{fig2}. Our domain-adapted depth estimation network achieved RMSE of 4.382 ± 1.304 mm, surpassing CycleGAN and network trained solely on virtual data, demonstrating the effectiveness of transfer learning for the network to learn domain common knowledge between synthetic and real data.

\section{Conclusions}
In this work, we introduce a synthetic-to-real domain adaptation framework for depth estimation in bronchoscopy. The network is initially trained on labeled synthetic data and then successfully transferred to real bronchoscopy data, demonstrating the viability of using synthetic data for depth estimation in bronchoscopy, potentially assisting in bronchoscopic procedures and other minimally invasive surgeries.

\addtolength{\textheight}{-4cm}   


\bibliographystyle{unsrt}
\bibliography{reference.bib}

\begin{thebibliography}{1}

\bibitem{teed2024deep}
Zachary Teed, Lahav Lipson, and Jia Deng.
\newblock Deep patch visual odometry.
\newblock {\em Advances in Neural Information Processing Systems}, 36, 2024.

\bibitem{bhat2023zoedepth}
Shariq~Farooq Bhat, Reiner Birkl, Diana Wofk, Peter Wonka, and Matthias M{\"u}ller.
\newblock Zoedepth: Zero-shot transfer by combining relative and metric depth.
\newblock {\em arXiv preprint arXiv:2302.12288}, 2023.

\bibitem{karaoglu2021adversarial}
Mert~Asim Karaoglu, Nikolas Brasch, Marijn Stollenga, Wolfgang Wein, Nassir Navab, Federico Tombari, and Alexander Ladikos.
\newblock Adversarial domain feature adaptation for bronchoscopic depth estimation.
\newblock In {\em Medical Image Computing and Computer Assisted Intervention--MICCAI 2021: 24th International Conference, Strasbourg, France, September 27--October 1, 2021, Proceedings, Part IV 24}, pages 300--310. Springer, 2021.

\bibitem{tian2024ddvnb}
Qingyao Tian, Huai Liao, Xinyan Huang, Jian Chen, Zihui Zhang, Bingyu Yang, Sebastien Ourselin, and Hongbin Liu.
\newblock Dd-vnb: A depth-based dual-loop framework for real-time visually navigated bronchoscopy.
\newblock {\em arXiv preprint arXiv:2403.01683}, 2024.

\bibitem{cyclegan}
Jun-Yan Zhu, Taesung Park, Phillip Isola, and Alexei~A Efros.
\newblock Unpaired image-to-image translation using cycle-consistent adversarial networks.
\newblock In {\em Proceedings of the IEEE international conference on computer vision}, pages 2223--2232, 2017.

\bibitem{bau2019seeing}
David Bau, Jun-Yan Zhu, Jonas Wulff, William Peebles, Hendrik Strobelt, Bolei Zhou, and Antonio Torralba.
\newblock Seeing what a gan cannot generate.
\newblock In {\em Proceedings of the IEEE/CVF international conference on computer vision}, pages 4502--4511, 2019.

\bibitem{ganin2016domain}
Yaroslav Ganin, Evgeniya Ustinova, Hana Ajakan, Pascal Germain, Hugo Larochelle, Fran{\c{c}}ois Laviolette, Mario March, and Victor Lempitsky.
\newblock Domain-adversarial training of neural networks.
\newblock {\em Journal of machine learning research}, 17(59):1--35, 2016.

\end{thebibliography}

\end{document}